\documentclass[aps,pret,a4paper,amsmath,twocolumn,floatfix]{revtex4}
\usepackage{graphics}
\usepackage{graphicx}
\usepackage{amsmath}
\usepackage{amssymb}
\usepackage{amscd}
\begin{document}

\title{Depletion effects in smectic phases of hard rod--hard sphere mixtures}

\author{Yuri Mart\'{\i}nez-Rat\'on}
\email{yuri@math.uc3m.es}

\affiliation{Grupo Interdisciplinar de Sistemas Complejos (GISC),
Departamento de Matem\'aticas, Escuela Polit\'ecnica Superior,
Universidad Carlos III de Madrid,
Avenida de la Universidad 30, E-28911 Legan\'es, Madrid, Spain.
}

\author{Giorgio Cinacchi}
\email{g.cinacchi@sns.it}

\affiliation{Dipartimento di Chimica, Universit\`{a} di Pisa
Via Risorgimento 35, I--56126, Pisa, ITALY}

\author{Enrique Velasco}
\email{enrique.velasco@uam.es}

\affiliation{Departamento de F\'{\i}sica Te\'orica de la Materia Condensada
and Instituto de Ciencia \\ de Materiales Nicol\'as Cabrera,
Universidad Aut\'onoma de Madrid, E-28049 Madrid, Spain.}

\author{Luis Mederos}
\email{l.mederos@icmm.csic.es}

\affiliation{Instituto de Ciencia de Materiales, Consejo Superior de
Investigaciones Cient\'{\i}ficas, E-28049 Cantoblanco, Madrid, Spain.}

\date{\today}

\begin{abstract}
It is known that when hard spheres are added to a pure system of hard rods
the stability of the smectic phase may be greatly enhanced, and that this 
effect can be rationalised in terms of depletion forces. In the present paper 
we first study the effect of orientational order on depletion forces in this
particular binary system, comparing  our results with those
obtained adopting the usual approximation of considering 
the rods parallel and their orientations frozen. We consider 
mixtures with rods of different aspect ratios and spheres of different diameters, 
and we treat them within Onsager theory.  Our results
indicate that depletion effects, and
consequently smectic stability, decrease significantly as a
result of orientational disorder in the smectic phase when compared with 
corresponding data based on the frozen--orientation approximation. These results
are discussed in terms of the $\tau$ parameter, which has been proposed as a 
convenient measure of depletion strength. We present closed expressions for $\tau$,
and show that it is intimately connected with the depletion potential. 
We then analyse the effect of particle
geometry by comparing results pertaining to systems of 
parallel rods of different shapes (spherocylinders, cylinders and parallelepipeds).
We finally provide results based on the Zwanzig approximation of a 
Fundamental--Measure density--functional theory 
applied to mixtures of parallelepipeds and cubes
of different sizes. In this case,  we show that the $\tau$ parameter
exhibits a linear asymptotic behaviour in the limit of large values of
the hard--rod aspect ratio, in conformity with Onsager theory, 
as well as in the limit of large values of the ratio of rod breadth to cube side length,  
$d$, in  contrast to Onsager approximation, which predicts $\tau\sim d^3$.
Based on both this result and the Percus--Yevick approximation for the direct 
correlation function for a hard sphere binary mixture in the same limit of 
infinite asymmetry, we speculate that, for spherocylinders and spheres, 
the $\tau$ parameter should be of order unity as $d$ tends to infinity. 
\end{abstract}


\maketitle

\section{Introduction }

In recent years experimental mixtures  that closely resemble a
hard rod--hard sphere  system have been studied; typically the rods 
are represented by tobacco mosaic or {\it fd}--virus  particles, while   
the spheres are represented by polystyrene latex particles or 
globular proteins \cite{fradenBPJ,fradenNATURE,fradenCOCIS}.
The phase diagrams found include the isotropic phase, as well as phases
with liquid-crystalline symmetry. Among these, nematic (N)
and smectic (Sm)
phases have been found. In addition, bulk
demixing transitions, as well as microsegregated phases of various
symmetries, have been observed\cite{fradenNATURE}. Intensive theoretical
effort has been devoted to the understanding of these systems. A useful
concept in this effort is that of a {\it depletion} force.
Structural and thermodynamic stability
of systems consisting of hard particles can be understood solely in terms of
entropic effects which, in mixtures, can be reinterpreted as an
attractive depletion force. Some studies have been done recently
in an attempt to quantify attractive depletion interactions between
solute particles, of anisotropic shape, mediated by solvent particles
that can be isotropic or possess themselves liquid-crystalline order
\cite{Cheung,MTdG,Roth}.

One of the microsegregated phases that has been observed is the lamellar
phase. This phase, having smectic symmetry,
consists of alternate pure layers of rods and spheres.
The lamellar phase can be greatly stabilised with respect to 
the corresponding smectic phase in the pure--rod fluid, as predicted in Ref. 
\cite{KodaJPSJ},
and shown in Refs. \cite{KodaMCLC,KodaPTP,DogicPRE,JPCM2004,LagoJMR,VeselyMP}.
In the lamellar phase, depletion forces result from the fact that it is
entropically more favourable for spheres to occupy the interstitial
space, which creates a large effectively attractive force between adjacent 
layers of rods; enhanced smectic--phase stability ensues. 

Thus far, theoretical studies of the lamellar phase have mostly used, among others, 
the following approximations: parallel rodlike particles with frozen orientational order, specific particle geometry, Onsager theory \cite{Onsager49} and neglect of particle 
flexibility. Cinacchi et al.\cite{Nosotros} studied the
hard-spherocylinder (HSPC)--hard-sphere (HS) mixture using Onsager theory and
free particle orientations, but presented phase diagrams including
isotropic, nematic and the lamellar phase, without specifically addressing
any stability issue. In this paper we lift stepwise the first three 
of these restrictions and assess their impact on
the enhancement of smectic stability in the mixture with respect to the
pure rod system.

Density--functional theory is a convenient theoretical tool to
study the structure and phase behaviour of soft--condensed matter.
Onsager theory is a most successful version of density--functional
theory, intended for bodies interacting through hard potentials.
Some of the shortcomings of Onsager theory are improved by
Fundamental--Measure theory \cite{Rosenfeld} (FMT), which 
introduces the fundamental particle measures in the theory and
should therefore be quite general. Our theoretical analysis 
is based on both these density--functional approximations.
Specifically, we first use the HSPC model within Onsager theory
to include the
orientational degrees of freedom of the rods by means of an
orientational distribution function and the associated order parameter. 
The results will be discussed in terms of a parameter, $\tau$, which
is directly related to the depletion strength in smectic phases \cite{DogicPRE}.
Then,  we discuss the effect of particle geometry by considering 
various shapes for the hard particles in the frozen-orientation
approximation.
Finally, we consider a FMT approximation for mixtures of hard 
parallelepipeds (HPAR) and cubes (HC),
and analyse it in the context of the Zwanzig approximation \cite{Yuri}; the latter 
consists of considering particle orientations to be restricted to three mutually 
perpendicular axes. Recent advances on the application of FMT to freely rotating 
particles have focused on anisotropic particles with infinitely narrow breadths 
(needles)\cite{Esztermann}. Since one of our aims is to investigate how
the depletion mechanism changes as the breadth ratio of the particles is
varied, these recent developments are not appropriate in our context, and
the use of the Zwanzig approximation seems to be justified for lack of a better approach.
FMT provides a different theoretical viewpoint with respect
to which the predictions of Onsager theory
can be assessed. In fact, the main purpose of this part of the work is to assess the 
impact of a proper inclusion in the theory of pair correlations
on the depletion mechanism in smectic phases.
In the framework of FMT, we additionally  study in detail
the asymptotic limit of $\tau$ with respect to different parameters relating to 
the aspect ratio of the rod and the relative sizes of HPARs and HCs.
With a view to completing this study, and based on some expressions
used to calculate the $\tau$ parameter, which we evaluate for hard spheres,
we speculate that, for HSPC--HS mixtures, $\tau\sim 1$ as the ratio
of rod breadth to sphere diameter tends to infinity.

In the following section the general procedure used to locate the nematic--smectic
spinodal line is illustrated; for the sake
of clarity,  this is illustrated in the context of Onsager theory for a binary mixture
of hard rods and HS's. 
In Section \ref{III} the $\tau$ parameter, which is used to quantify the
stability of the smectic phase when HS's are added, is defined, and a 
closed expression, valid in the context of   Onsager theory and parallel rods, is presented.
The results are  presented in Section \ref{V}, 
which has three subsections. The first is devoted to the effect of
free particle orientations, the second addresses the effect of
particle shape and the third  contains 
the  results for mixtures of HPAR's and HC's,
analysed by means of FMT in the Zwanzig approximation. This last 
subsection is in turn divided into two parts. The first is devoted to the 
results obtained by applying Onsager theory to the Zwanzig  
HPAR--HC system, while in the second we describe the results from FMT,
together with a detailed discussion on the asymptotic limits of 
the $\tau$ parameter. 
The conclusions are presented in Section \ref{VI}. The Appendices collect
expressions for the Fourier transforms of the overlap functions
for the various particle geometries explored in the paper, together
with a few details on the numerical minimisation
of the functional in the context of the frozen--orientation approximation.
An exact expression for $\tau$ in terms of the correlation functions,
valid for a general mixture of freely rotating particles, is 
also included. Explicit expressions for these functions in the
Zwanzig--Onsager and Zwanzig--FMT approaches for HPAR--HC and HS binary mixtures
are provided. Finally, the asymptotic behaviour of the $\tau$ 
parameter, within the framework of FMT, and for the HPAR--HC mixture,
is described.

\section{Onsager theory for hard rod--hard sphere mixtures and stability analysis}
\label{II}

The first theory we employ to describe hard rod--hard sphere mixtures is 
Onsager theory, first proposed by Onsager\cite{Onsager49} for a pure system
of hard rods and extensively used in theoretical approaches to orientational
ordering in hard--rod fluids. It can be regarded as a truncated virial expansion, up to
second order in density, of the excess part of the free energy. Thus, it contains
the orientational--dependent second--virial coefficient exactly. It is a
density--functional theory, since the free energy depends functionally on the
orientational distribution function. The extension of Onsager theory to
mixtures is straightforward; the reader is referred to Ref.\cite{Nosotros} for
a detailed account on its implementation for the present model fluids.

We consider a two-component mixture of hard rods and hard spheres
labelled with 1 and 2, respectively. The density functional in the
Onsager approximation, assuming uniaxial smectic symmetry and taking $z$ as
the coordinate along the smectic layer normal, is written as
\begin{eqnarray}
\Phi&=&\frac{1}{d_s}\int_0^{d_s} dz \left\{
\sum_{i=1}^2\rho_i(z)\left[\log{\left(\frac{\rho_i(z)}{4\pi}\right)}-1\right]
-\rho_1(z)S_{\hbox{\tiny r}}(Q)
\right.\nonumber\\&&\left.\right.\nonumber\\
&+&\left.\frac{1}{2}\sum_{i,j=1}^2
\int dz^{\prime}\rho_i(z)\rho_j(z^{\prime}) f_{ij}(z-z^{\prime};Q)\right\},
\label{ONS}
\end{eqnarray}
where $\Phi\equiv\beta{\cal F}/V$ is the Helmholtz free-energy density per
unit thermal energy, $d_s$ the smectic layer spacing,
$S_{\hbox{\tiny r}}(Q)$ is the orientational-entropy
density of the hard rods, $Q$ their orientational order parameter, 
$\rho_1(z)$ and $\rho_2(z)$ are local densities for the two components,
and $f_{ij}(z)$ are angular averages of the overlap functions
between species $i$ and $j$, already integrated 
in the $xy$ plane. The procedure used to calculate these averages
has been described in detail elsewhere 
for the case of mixtures of spherocylinders \cite{Nosotros}. In the case where one adopts
the popular approximation of considering that particles possess perfect
orientational order ($Q=1$), an approximation used in  all 
previous theoretical analyses of the hard-rod--hard-sphere mixture 
\cite{KodaJPSJ,KodaMCLC,KodaPTP,DogicPRE,VeselyMP},
the functions $f_{ij}(z)$ can usually be written exactly,
depending on the particle geometry.
Note that $f_{11}$ and $f_{12}$ both depend on $Q$, 
but certainly not $f_{22}$.
In the nematic phase, $\rho_i(z)$ is a constant.

Since we are interested in searching for the nematic-smectic spinodal, 
we consider the following perturbations on the 
constant nematic densities:
\begin{eqnarray}
\rho_i(z)=\rho x_i\left(1+\lambda_i\cos{kz}\right),\hspace{0.6cm}
\hspace{0.6cm}i=1,2
\end{eqnarray}
where $\rho$ is the total mean density, $x_2\equiv x$, $x_1=1-x$ 
are the molar fractions of the two components, and $k$ is the smectic wave 
number. 
Note that $\lambda_i$
can be positive or negative; in the case that $\lambda_1\lambda_2<0$, 
smectic layers rich in one of the components will alternately appear,
giving rise to a microsegregated smectic phase. 
The free energy difference between smectic and nematic states, to
second order in the $\lambda_i$'s, is:
\begin{eqnarray}
&&\frac{\beta\Delta F}{N}=\frac{1}{4}\left\{\sum_{i=1}^2 x_i\lambda_i^2
+\rho\sum_{i,j=1}^2x_ix_j\lambda_i\lambda_j\tilde{f}_{ij}(k;Q)\right\}
\end{eqnarray}
where $\tilde{f}_{ij}(k)$ are the following cosine Fourier transforms:
\begin{eqnarray}
\tilde{f}_{ij}(k)=\int_{-\infty}^{\infty} dt f_{ij}(t)\cos{kt},
\hspace{0.6cm}i,j=1,2.
\end{eqnarray}
The Hessian matrix,
$\partial^2(\beta\Delta F/N)/\partial\lambda_i\partial\lambda_j$, 
necessary to study the stability of the nematic
phase against smectic fluctuations, is
\begin{eqnarray}
H=\left(\begin{array}{cc}
\displaystyle\frac{x_1}{2}\left[1+\rho x_1\tilde{f}_{11}(k;Q)\right]&
\displaystyle\frac{\rho x_1x_2}{2}\tilde{f}_{12}(k;Q)\\\\
\displaystyle\frac{\rho x_1x_2}{2}\tilde{f}_{12}(k;Q)&
\displaystyle\frac{x_2}{2}\left[1+\rho x_2\tilde{f}_{22}(k)\right]
\end{array}\right)
\end{eqnarray}
The spinodal line is then obtained by solving the equations
\begin{eqnarray}
&&\Delta(k_0,\rho_0,Q_0)\equiv\left[1+\rho_0 x_1\tilde{f}_{11}(k_0;Q_0)\right]
\left[1+\rho_0 x_2\tilde{f}_{22}(k_0)\right]\nonumber\\&&\nonumber\\&&-
\rho_0^2x_1x_2\left[\tilde{f}_{12}(k_0;Q_0)\right]^2=0\nonumber\\&&\nonumber\\
&&\left.\frac{\partial\Delta}{\partial k}\right|_{\rho_0,k_0,Q_0}=0,\hspace{0.6cm}
\left.\frac{\partial F}{\partial Q}\right|_{\rho_0,k_0,Q_0}=0
\label{ec}
\end{eqnarray}
for $\rho_0$, $k_0$ and $Q_0$, the  values of the density, wave-vector
and order parameter of the unstable nematic at the spinodal, respectively;
note that these quantities will depend on the value of the composition $x$.
The second equation ensures that instability will occur at some 
particular $k$ vector for the first time, as the density is 
decreased from above down to the value at the spinodal.
For the pure case ($x=0$) the instability equations reduce to
\begin{eqnarray}
&&1+\rho_0\tilde{f}_{11}(k_0;Q_0)=0,\\&&\nonumber\\
&&\left.\frac{\partial\tilde{f}_{11}}{\partial k}\right|_{\rho_0,k_0,Q_0}=0,\hspace{0.6cm}
\left.\frac{\partial F}{\partial Q}\right|_{\rho_0,k_0,Q_0}=0.
\label{ec0a}
\end{eqnarray}
These are the equations defining a spinodal to an ordered phase in a pure system:
$1-\rho_0 \tilde{c}(k_0)=0$, where $\tilde{c}(k)$ is the Fourier transform of
the direct correlation function (in Onsager theory $\tilde{c}(k)=-\tilde{f}(k)$).

\section{The $\tau$ parameter}
\label{III}
The stability of the nematic phase against smectic-type fluctuations
can be quantified in different ways. In this paper we use the parameter $\tau$
proposed by Dogic et al.\cite{DogicPRE} in their 
analysis of the hard-rod-hard-sphere mixture:
\begin{equation}
\tau=\left.\lim_{\eta_2\to 0}\frac{d\eta}{d\eta_2}\right|_{\hbox{\tiny 0}}.
\end{equation}
In this expression $\eta$ is the total packing fraction of the mixture,
$\eta_2$ the partial packing fraction of spheres, and the derivative is
evaluated at the nematic-smectic spinodal line. Other definitions
are possible; for example, one could use the spinodal line $p(\eta_2)$,
with $p$ the pressure, instead of $\eta(\eta_2)$. Here we adhere to
the $\tau$ parameter as a measure of smectic stability. It turns out,
as shown below, that the $\tau$ parameter can be directly related to
the depletion potential, so that $\tau$ contains basic information on
depletion forces in the system.

We now obtain a compact expression for $\tau$ in the framework of 
Onsager theory, considering the case of perfect orientational order ($Q=1$). The
generalization to free orientational order and a general density
functional is given in the Appendix C. Our
strategy consists of obtaining the nematic--smectic spinodal line $\eta=\eta(\eta_2)$
perturbatively in the packing fraction $\eta_2$ (i.e. for small values of $\eta_2$), and 
then extracting the value of $\tau$ from the first-order term. We begin by first
expanding Eqns. (\ref{ec}) at small $x$. We assume the expansions
\begin{equation}
\rho=\rho_0+\rho_0^{\prime} x+...,\hspace{0.6cm}
k=k_0+k_0^{\prime} x+...,\hspace{0.6cm}
\end{equation}
where now $\rho_0$ and $k_0$ are the density and wave-vector, at the $x=0$ spinodal,
and $\rho_0^{\prime}$ and $k_0^{\prime}$ are the derivatives of $\rho$ and $k$
with respect to the molar fraction $x$,  at the $x=0$ spinodal, respectively. 
Inserting these
equations into the first two of Eqns. (\ref{ec}) we obtain, at zeroth order in $x$:
\begin{equation}
1+\rho_0\tilde{f}_{11}(k_0)=0,\hspace{0.6cm}
\tilde{f}_{11}^{\prime}(k_0)=0.
\label{14}
\end{equation}
To first order in $x$ we get
\begin{equation}
\frac{\rho_0^{\prime}}{\rho_0}=1-\rho_0^2\tilde{f}_{12}^2(k_0),\hspace{0.6cm}
k^{\prime}=-2\frac{\tilde{f}_{12}(k_0)\tilde{f}_{12}^{\prime}(k_0)}
{\tilde{f}_{11}(k_0)\tilde{f}_{11}^{\prime\prime}(k_0)}.
\end{equation}
Now, differentiating the equations $\eta=\eta_1+\eta_2=\rho(1-x)v_1+\eta_2$ and
$\eta_2=\rho x v_2$, and taking the limit $x\to 0$, we obtain the relation
\begin{equation}
\tau=\lim_{x\to 0}\frac{d\eta}{d\eta_2}=
1-\frac{v_1}{v_2}\left(1-\frac{\rho_0^{\prime}}{\rho_0},
\label{relation}
\right)
\end{equation}
where $v_1$, $v_2$ are the volumes of the two particle species, rods and
spheres respectively. This equation, along with Eqn. (\ref{14}), gives 
$\eta_0=-v_1/\tilde{f}_{11}(k_0)$, and
\begin{equation}
\tau=1-\eta_0^2\frac{\tilde{f}_{12}^2(k_0)}{v_1v_2}=1-
\frac{v_1}{v_2}\left[\frac{\tilde{f}_{12}(k_0)}{
\tilde{f}_{11}(k_0)}\right]^2.
\label{tau}
\end{equation}
In fact, this expression remains valid for binary mixtures composed of
any convex bodies, but can only be used in the approximation of frozen particle 
orientations ($Q=1$) and in the framework of the Onsager approximation.
The $\tau$ parameter is calculated by solving Eqns. (\ref{14}) and then
evaluating Eqn. (\ref{tau}).

We now proceed to show the direct relation existing between the 
parameter $\tau$ and the depletion interaction
between two parallel rods mediated by the solvent 
particles (hard spheres), at least in the limit where the density
of the solvent particles $\rho_2$ and their diameter, relative to the breadth
of the rods, are very small. These are the conditions under which 
the Asakura--Oosawa
approximation \cite{AO} is valid, and the depletion potential becomes
$V_{\rm{dep}}({\bf r},\rho_2)=-\rho_2 \left[f_{12}\ast f_{12}\right]({\bf r})$
(the asterisk standing for convolution). Now
\begin{eqnarray}
\frac{\partial\tilde{V}_{\rm{dep}}(k_0)}{\partial\rho_2}=-\tilde{f}_{12}^2(k_0),
\label{Vdep}
\end{eqnarray}
and Eqn. (\ref{tau}) gives
\begin{equation}
\tau=1+\frac{v_1}{v_2}\frac{\displaystyle{\frac{\partial \tilde{V}_{\rm{dep}}}
{\partial \rho_2}(k_0)}}{\tilde{f}^2_{11}(k_0)}
\label{imply}
\end{equation}
(this equation can also be obtained from a mean--field perturbative
treatment of the fluid, by considering parallel hard rods that interact 
via a depletion potential in mean--field approximation).
Eqn. (\ref{imply}) relates the $\tau$ coefficient with the depletion 
potential, and shows that $\tau$ contains basic information on 
depletion forces. In turn, it provides a condition under which 
$\tau$ may be negative, and thus directly links the depletion potential
to the enhancement of smectic--phase stability in fluid mixtures of
hard rods and spheres. 

\section{Results}
\label{V}

\subsection{Mixtures of HSPC and HS: frozen versus free orientations}

\begin{figure}[h]
{\centering\resizebox{0.45\textwidth}{!}{\includegraphics{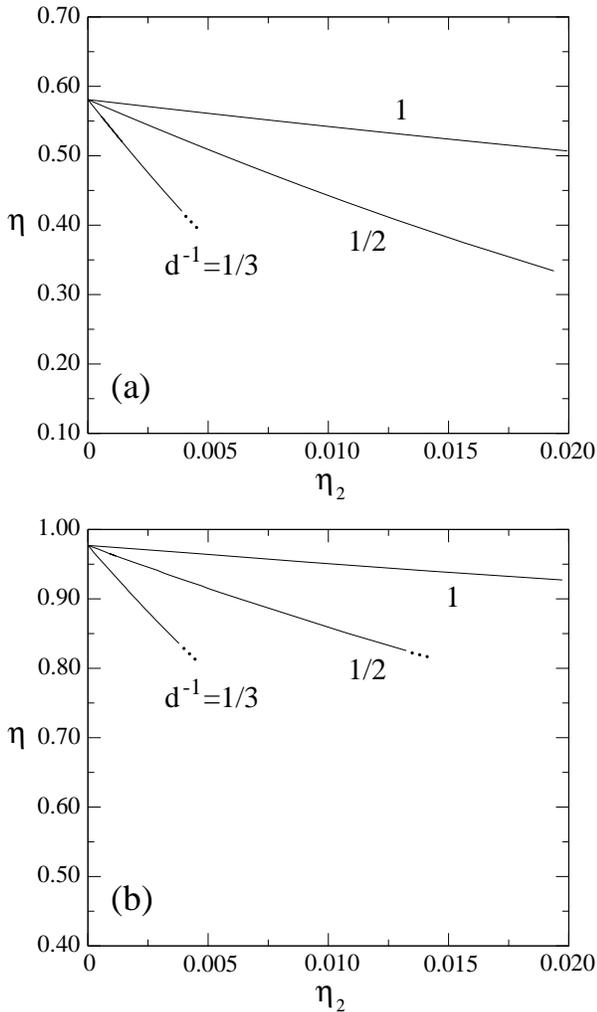}} }
\caption{\label{fig1}\small Nematic--smectic spinodal line $\eta(\eta_2)$,
according to Onsager theory, for the case $\kappa_1=8$, and $d^{-1}=1$, $1/2$ 
and $1/3$ (indicated as labels). (a) Frozen--orientation approximation; 
(b) free orientations. Note that horizontal and vertical scales in both graphs
are the same.} 
\end{figure}

\begin{figure}[h]
{\centering \resizebox{0.45\textwidth}{!}{\includegraphics{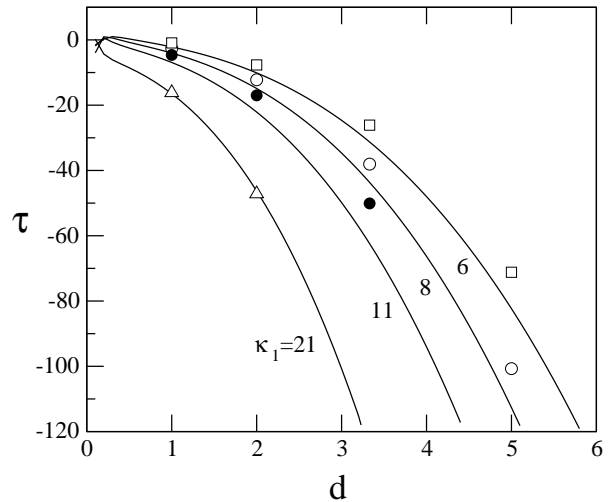}} \par}
\caption{\label{taus}\small
Values of $\tau$ as a function of inverse HS diameter ratio $d$,
for various values of HSPC length--to--breadth ratio $\kappa_1$. Lines: Frozen--orientation
approximation 
 ($Q=0$); from top to bottom: 
$\kappa_1=6$, $8$, $11$ and $21$. Symbols: free  
orientations. Squares: $\kappa_1=5$; open circles: $\kappa_1=8$; filled
circles: $\kappa_1=11$, and triangles: $\kappa_1=21$.}
\end{figure}

All previous theoretical analyses of the lamellar phase in 
the hard-rod--hard-sphere mixtures 
\cite{KodaJPSJ,DogicPRE,VeselyMP} 
have relied on the approximation of considering perfect orientational order.
In this section we assess the impact of this severe approximation on the
formation of smectic phases in the HSPC/HS mixture.
We have investigated a number of HSPC/HS mixtures using the original
Onsager theory, changing the HSPC aspect ratio $\kappa_1\equiv(L_1+D_1)/D_1$ 
and the HSPC breadth relative to the HS diameter, $d\equiv D_1/D_2$. 
Fig. \ref{fig1} 
shows the spinodal line $\eta$--$\eta_2$ of the nematic--smectic transition,
for the cases of frozen and free orientations, and for $\kappa_1=8$ and
various values of the scaled HS diameter. We note that, in all cases shown,
the smectic phase is stabilised with respect to the nematic phase (negative
slope at $\eta_2=0$) on adding HS to the pure HSPC fluid; this is always the case 
when $d\ge 1$ (spheres smaller than the cylinder diameter). The effect is
amplified when both the HSPC aspect ratio length ($\kappa_1$) is increased and 
the sphere diameter is decreased ($d$ increases). However, the effect 
is less pronounced when the HSPC are free to orient their main axes. The $\tau$ 
parameter, shown in Fig. (\ref{taus}), reflects this behaviour, which points to
a less strong depletion effect due to orientational fluctuations.
\subsection{Other hard rod--hard sphere mixtures: effect of particle shape}
\label{shapes}
To investigate the effect of the shape of the rods on the depletion 
effect and, in turn, on the stability of the smectic phase, Vesely\cite{VeselyMP}
has recently considered parallel-rod models consisting of linear overlapping 
hard-sphere chains, hard ellipsoids and hard spheroellipsoids.
Here we have studied, still in the approximation of frozen orientations, two
additional types of hard particles: cylinders (HCYL) and parallelepipeds with a
square transverse shape (HPAR).
To minimize the effect of the different particle geometries,
the particles were chosen to have the same volume (see later).
In all cases 
the minority component continues to consist of hard spheres.
The functions $f_{ij}(z)$, and also their Fourier transforms,
can be calculated analytically in these cases. The relevant expressions
can be found in the Appendix A. 
\begin{figure}[h]
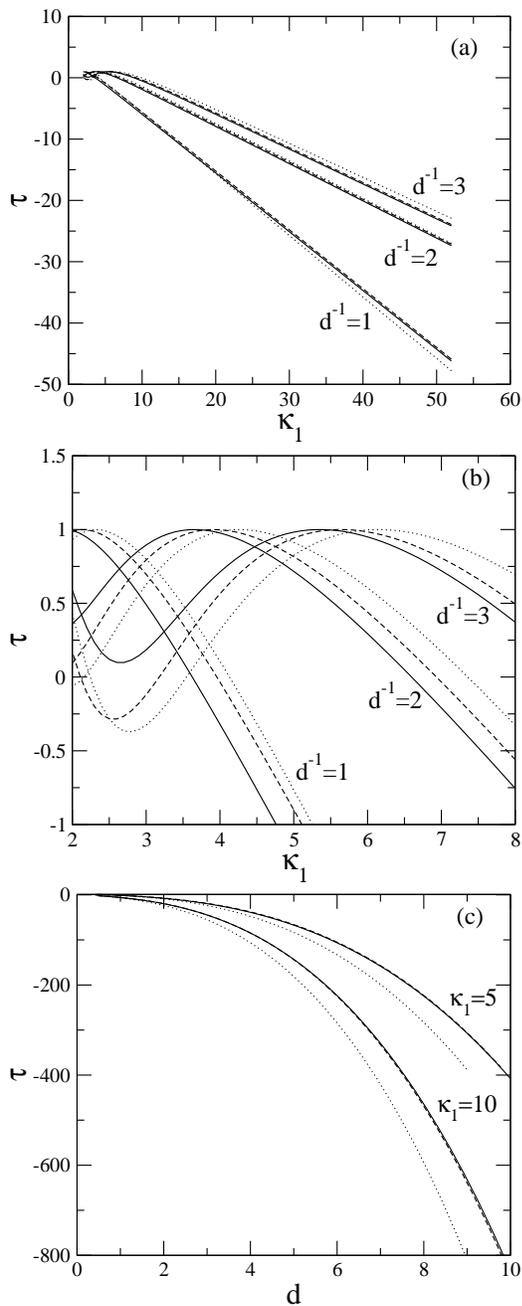

{\centering \resizebox{0.38\textwidth}{!}{\includegraphics{fig3a.eps}} \par}
{\centering \resizebox{0.38\textwidth}{!}{\includegraphics{fig3b.eps}} \par}
{\centering \resizebox{0.38\textwidth}{!}{\includegraphics{fig3c.eps}} \par}
\caption{\small
(a) The $\tau$ parameter vs. $\kappa_1$ for different values of the 
parameter $d$ (indicated in the graph). (b) A zoom of 
(a) for small $\kappa_1$'s. (c) $\tau$ vs. $d$ for different values of 
$\kappa_1$ as indicated in the figure. (c) $\tau$ parameter as a function 
of $d$ for different values of $\kappa_1$, as indicated. 
Results for HSPC, HCYL, and HPAR are plotted with 
solid, dashed, and dotted lines, respectively.}
\label{final}
\end{figure}

Results for the $\tau$ parameter in the case of HCYL/HS and HPAR/HS mixtures, 
along with the previously considered HSPC/HS mixtures, are shown in Figs.
\ref{final}(a)-(c). The $\tau$ parameter is plotted against the
aspect ratio of the rods $\kappa_1$ in Fig. \ref{final}(a)
for different values of the scaled inverse HS diameter $d=D_1/D_2$ 
($D_1$ being the side length of the parallelepiped in the case of HPAR);
Fig. \ref{final}(b) is a zoom of (a) in the region of small $\kappa_1$. Finally, the 
$\tau$ parameter against $d$ for different values $\kappa_1$ is plotted in 
Fig. \ref{final}(c). 
We can see from Fig. \ref{final}(a) that  $\tau$ tends to decrease
linearly, becoming negative for sufficiently large aspect ratios.
This effect is more and more pronounced as the diameter of the HS
is decreased. An interesting feature is that,
when the diameter of HS is greater than that of the rod, depletion is larger, 
and therefore smectic stability is enhanced more substantially,
as one goes from HPAR to HCYL or HSPC (note that these two are very similar). 
The opposite behaviour (i.e. 
depletion in HPAR is enhanced with respect to HCYL and HSPC) results
when the HS diameter is equal or less than that of the HSPC [Fig. 
\ref{final}(a)]. For short rods $\tau$ exhibits oscillatory behaviour 
[see Fig. \ref{final}(b)]. Finally, 
fixing $\kappa_1$ and increasing $d$, we find that the $\tau$ parameter
decreases as a cubic power-law in $d$ 
and that its value for HPAR is much less than that for HSPC or HCYL.

The linear and cubic dependences of $\tau$ with respect to $\kappa_1$ and 
$d$, respectively, can be elucidated from Eqn. (\ref{tau}) using the fact
that the volume ratio can be expressed as $v_1/v_2=\kappa_1 d^3$ for the HCYL
case (the other cases being similar), while the squared ratio of Fourier 
transforms of the overlap functions tends asymptotically, for large $\kappa_1$  
and $d$, to a constant. Simple calculations give, in this limit,
\begin{eqnarray}
\tau_{\hbox{\tiny HC}}=
\tau_{\hbox{\tiny HSPC}}=-\frac{3}{128}\left(\cos{\frac{k_0^*}{2}}\right)^{-2}
\kappa_1 d^3
\end{eqnarray}  
with $k_0^*=k_0L_1$ the reduced wave number of the N--Sm spinodal 
instability in the one--component system. 

\begin{figure}[h]
{\centering \resizebox{0.45\textwidth}{!}{\includegraphics{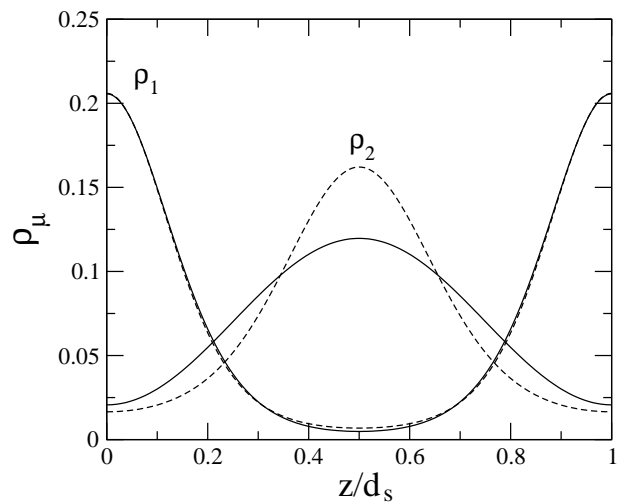}} \par}
\caption{\small Equilibrium density profiles of the equimolar HSPC--HS 
mixture (different components are labelled in the figure) 
resulting from minimization with respect to the Fourier 
amplitudes (solid lines) and with respect to a single--amplitude exponential
parametrization (dashed lines). The HSPC has an aspect ratio $\kappa_1=10$ and 
its breadth coincides with the HS diameter, taken as unity. The 
pressure in reduced units was chosen to be $0.4$.}  
\label{diff}
\end{figure}

To study the effect of a varying particle shape on the 
phase behaviour of the rod-sphere mixtures, we have performed 
a full minimization 
of the Helmholtz free-energy functional with respect to smectic-like density 
profiles of the rods [$\rho_1(z)$] and spheres [$\rho_2(z)$].  
The minimization was carried out with respect to the Fourier amplitudes of 
the truncated Fourier expansion of the density profiles 
and smectic wave number (details are provided in  Appendix B). 
The choice of a Fourier expansion is justified by the fact that the
HS equilibrium density profiles obtained with this minimization differ very
significantly from those obtained via the usual single--amplitude exponential
parametrization, as shown in 
Fig. \ref{diff}. We can see that the parametrization produces too sharp 
density peaks in the HS profile, which enhances the depletion effect. 

Since we would like to compare the results from the 
density--functional minimization of mixtures composed of rods 
of different geometries (HSPC, HCYL and HPAR), a sensible criterion
is required to choose the sizes of the particles so as to make sure 
that these differences arise from the shape and not from different 
sizes and volumes. For this purpose we have used three different 
criteria: (i) all particles have the same diameter and the 
same length, (ii) all particles have the same length and the 
same volume, and (iii) all particles have the same aspect ratio and 
the same volume. Comparing the equilibrium density profiles resulting 
from use of these three criteria, we have concluded that differences
between particles are minimised using the third criterion
(these results are not shown here).   

We have calculated, using the free, i.e. Fourier--based, minimization, the phase diagrams 
of binary mixtures of rods (HSPC, HCYL and HPAR) and HS. The HSPC and HS diameters
were taken to be equal and adopted as unit of length. The aspect ratios  
of all the rods were set to 10, all particle volumes being the same.
The following relations result:
\begin{eqnarray}
&&D_{\hbox{\tiny HC}}=
\left[1-\left(3\kappa_{\hbox{\tiny HSPC}}\right)^{-1}\right]^{-1/3} 
D_{\hbox{\tiny HSPC}}, \nonumber\\\nonumber\\&&
D_{\hbox{\tiny HP}}=\left(\frac{\pi}{4}\right)^{1/3}D_{\hbox{\tiny HC}},
\quad L_{\mu}=\kappa_{\hbox{\tiny HSPC}} D_{\mu},
\end{eqnarray} 
with $\mu$=HSPC, HC or HP.
\begin{figure}[h]
{\centering \resizebox{0.45\textwidth}{!}{\includegraphics{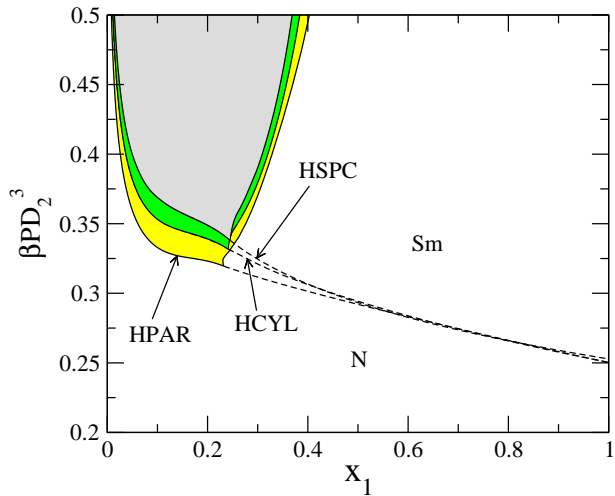}} \par}
\caption{\small Phase diagrams in the reduced pressure--composition plane 
of three different mixtures composed of rods of different 
geometries (HSPC, HCYL, and HPAR) with  HS.  Different lines corresponding 
to each mixture are labelled in the figure. Also indicated are the stability 
regions of the nematic and the smectic phases.  
Solid lines represent the binodals of the N--Sm coexistence, while the continuous
N--Sm transitions have been indicated by a dashed line.}  
\label{PHD}
\end{figure}
The resulting phase diagrams are shown in Fig. \ref{PHD}.
The first interesting feature of all  
phase diagrams is the presence of a 
tricritical point, where the nature of the N--Sm transition changes from 
second to first order\cite{tricritical}. 
For pressures higher than that of the tricritical point,
the demixing instability region becomes wider. The width of this 
demixing window increases for mixtures going from HSPC to HCYL and then HPAR.
This scenario clearly demonstrates that increasing differences  
in shape between spheres and rods enhance the depletion in smectics,
resulting in a wider demixing gap. 
   
\subsection{Zwanzig model: Onsager theory versus FMT}

In order to check whether the effects presented above are robust
with respect to the theory and approximations used, 
we have analysed a mixture of HPAR and HC using Fundamental--Measure
theory in the Zwanzig approximation. 
Zwanzig \cite{Zwanzig} introduced a model that considerably reduces the
complexities associated with minimising a density-functional
that depends on particle orientations. The approximation
consists of restricting the possible orientations of a rod to
lie along the three orthogonal axes $x,y$ and $z$, treating a pure
system as a three-component mixture (one component for each orientation). 
The approximation is very useful and has been used quite substantially.
In our FMT context, the resulting theory treats spatial correlations
very accurately so that it should better describe the
onset of smectic order in the nematic fluid.  
We have used the  HPAR-HC system and the Zwanzig approximation
because in this combination a FMT can be rigorously formulated \cite{Yuri}.
The Zwanzig  approximation
should not be terribly limiting, since the nematic fluid is well oriented.
In order to elucidate the effect of restricted orientations
separately from that inherent to the proper inclusion of higher--order correlations,
a study of the HPAR-HC mixture using a Zwanzig--Onsager theory,
together with that on mixtures of parallel HPAR and HC, is  necessary.  

\subsubsection{Zwanzig model in the Onsager formulation}

\begin{figure}[h]
{\centering\resizebox{0.45\textwidth}{!}{\includegraphics{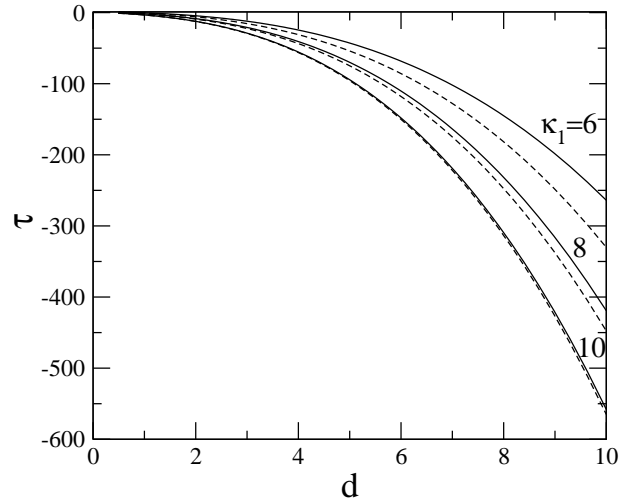}} \par}
\caption{\small $\tau$ parameter as a function of $d$ for three different 
values of $\kappa_1$, as indicated. Solid and dashed lines correspond to
the Zwanzig and parallel approaches, respectively.} 
\label{fig_Zwan}
\end{figure}

In Appendix D we have written expressions for the Fourier 
transforms of the correlations functions $\tilde{c}_{ij}(k)$ 
as obtained from the Zwanzig--Onsager approach.
These expressions are necessary
to calculate the $\tau$ parameter using Eqn. (\ref{ultima}). The 
results obtained are plotted in Fig. \ref{fig_Zwan}. 
In the same figure the parallel case is also plotted for the sake of comparison. 
The usual cubic power--law behaviour of $\tau$ with 
respect to $d$ is obtained;  this is typical of any Onsager approach,  as
discussed  in Sec. \ref{shapes}. The Zwanzig approach gives a 
lower value for $\tau$ as compared to the parallel case. 
This result is similar to that already obtained using the freely rotating 
Onsager approach for HSPC. For high values of $\kappa_1$, both approaches 
collapse into a single line, which is due to the 
high value of the order parameter $Q$ of a fluid composed of long 
particles.  

\subsubsection{Zwanzig model in the FMT formulation}
Details on this approximation can be found in Ref. \cite{Yuri}. Here we 
only give a brief sketch of the theory.  
We continue to use the notation introduced in the 
previous section for the dimensions of the particles, i.e. the length of
the HPAR is $L_1$, the side length is $D_1$ (these particles are assumed to have
a square section), and $D_2$ is the side length of the cubes.
Since the unit vector $\hat{\bf\Omega}$ only has three possible
orientations, the one--particle distribution functions 
$\rho_s({\bf r},\hat{\bf\Omega})$ can be expressed as
\begin{eqnarray}
\rho_s({\bf r},\hat{\bf\Omega})=
\sum_{\mu}\rho_{s\mu}({\bf r})\delta(\hat{\bf\Omega}-\hat{\bf e}_{\mu})
\label{densities}
\end{eqnarray}
where $\hat{\bf e}_{\mu}$, $\mu=1,2,3$, are unit vectors along the three 
perpendicular directions $xyz$, respectively, and
$\rho_{s\mu}({\bf r})$ are the local density of 
species $s$ parallel to the $\mu$-axis. 
The excess part of the free-energy density, in reduced thermal units,
is obtained in \cite{Yuri}, and has the form 
\begin{eqnarray}
\Phi_{\hbox{\tiny ex}}({\bf r};\{\rho_s\})=
-n_0\ln(1-n_3)+\frac{{\bf n}_1\cdot{\bf n}_2}{
1-n_3}+\frac{n_{2x}n_{2y}n_{2z}}{(1-n_3)^2},\nonumber\\
\label{7}
\end{eqnarray}
with the functions $\{n_{\alpha}\}$ ($\alpha=\{0,1x,1y,1z,2x,2y,2z,3\}$)
being weighted densities obtained as 
\begin{eqnarray}
n_{\alpha}({\bf r})=\sum_{s=1}^2\sum_{\mu=1}^3
\int_V d{\bf r}^{\prime}\rho_{s\mu}({\bf r}^{\prime})\omega_{s\mu}^{(\alpha)}
({\bf r}-{\bf r}^{\prime}),
\label{8}
\end{eqnarray}
where $\omega_{s\mu}^{(\alpha)}$ are characteristic functions whose
spatial integrals give the fundamental measures of the particles
(edge length, surface and volume). 
The ideal part of the free energy density in reduced thermal units 
for this model is 
\begin{eqnarray}
\Phi_{\rm{id}}(z)=\sum_{s\mu}\rho_{s\mu}(z)
\left[\log\rho_{s\mu}(z)-1\right],
\end{eqnarray}
so that the total free energy per unit volume and unit thermal energy can be 
calculated as 
\begin{eqnarray}
\Phi=\frac{1}{d_s}\int_0^{d_s} dz\left[\Phi_{\rm{id}}(z)+\Phi_{\rm{exc}}(z)
\right].
\label{phi}
\end{eqnarray}
Density and order--parameter profiles can be defined; in particular,
the density profile is
\begin{eqnarray}
\rho_s(z)=\sum_{\mu}\rho_{s\mu}(z).
\end{eqnarray}
The nematic--smectic spinodal line can be obtained in this theory 
using the same kind of arguments used for the Onsager theory (Sec. \ref{II}). Note, however,
that the expression (\ref{tau}) for the $\tau$ parameter is not
valid in this context, and we need to use the more general formula 
(\ref{ultima}), valid for any mixture of freely rotating particles. In 
the Appendix C we describe in detail how this formula
is obtained. An alternative way to calculate this parameter is 
to implement a numerical differentiation scheme on the
spinodal line $\eta(\eta_2)$.

\begin{figure}[h]
{\centering \resizebox{0.45\textwidth}{!}{\includegraphics{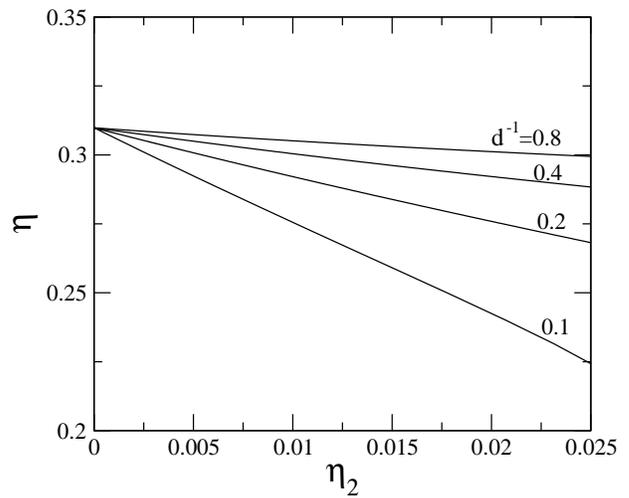}} \par}
\caption{\small
Nematic--smectic spinodal line $\eta$--$\eta_2$,
as obtained from Fundamental--Measure theory for the HPAR-HC mixture,
for $\kappa_1=8$ and various values of $d^{-1}$, as indicated
in the graph.}
\label{spinodalZwanzig}
\end{figure}

\begin{figure}[h]
{\centering \resizebox{0.45\textwidth}{!}{\includegraphics{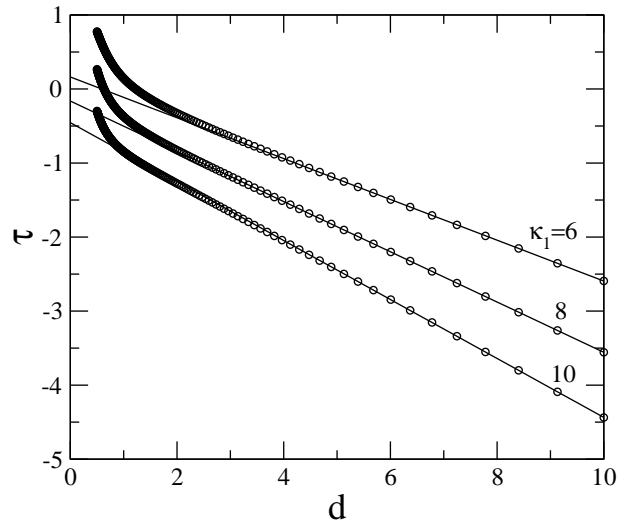}} \par}
\caption{\small
$\tau$ parameter as a function of $d$ for various values of
the parallelepiped length--to--breadth $\kappa_1=L_1/D_1$ (indicated in the graph). 
The data are indicated by symbols;  linear 
least-square fits $\tau=\alpha(\kappa_1)+\beta(\kappa_1)d$ 
are included as lines for each value of $\kappa_1$.}
\label{tauZwanzig1}
\end{figure}

\begin{figure}[h]
{\centering \resizebox{0.45\textwidth}{!}{\includegraphics{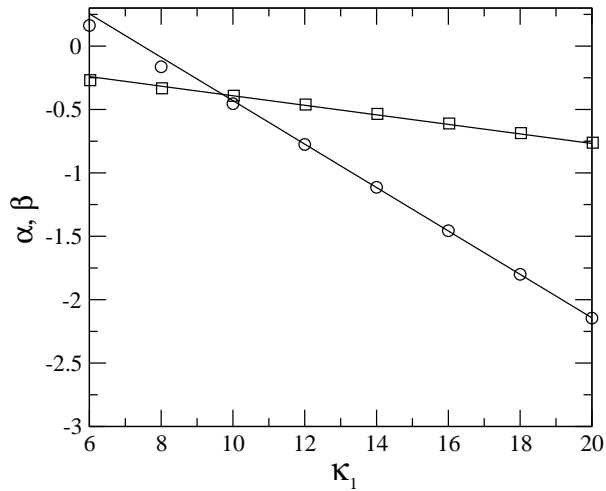}} \par}
\caption{\small
Symbols: coefficients $\alpha$ (circles) and $\beta$ (squares) as
a function of $\kappa_1$. Solid
lines: linear fits $\alpha(\kappa_1)=
\alpha_1+\alpha_2\kappa_1$ and $\beta(\kappa_1)=\beta_1+\beta_2\kappa_1$, 
with $\beta_2=-0.038$.}
\label{tauZwanzig}
\end{figure}

The nematic--smectic spinodal lines for the case
$\kappa_1=8$ and various values of the scaled cube side length $d^{-1}=D_2/D_1$ 
are shown in Fig. \ref{spinodalZwanzig}, whereas the corresponding $\tau$ parameters for
different values of $\kappa_1$ are shown in Fig. \ref{tauZwanzig1}. As was 
the case in the mixtures analysed 
using Onsager theory, $\tau$ decreases as the size of the cubes is diminished,
indicating a stronger depletion effect and a corresponding
enhancement of the smectic--phase stability.
An interesting feature of these results is that they appear to
exhibit linear behaviour for large values of $d$ (i.e. as
the size of the cubes becomes smaller). Fig. \ref{tauZwanzig1}
shows this by means of least--square fits to linear functions.
The coefficients of the fit are functions of the aspect ratio
$\kappa_1$, and,   in turn, 
appear to be linear in $\kappa_1$, as shown in Fig. \ref{tauZwanzig}. 

\begin{figure}[h]
{\centering \resizebox{0.45\textwidth}{!}{\includegraphics{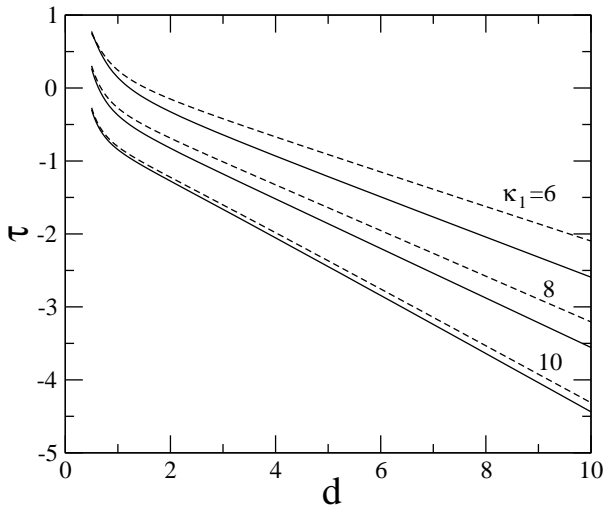}} \par}
\caption{\small The $\tau$ parameter as a function of $d$ 
calculated from the Zwanzig (solid line) and 
parallel (dashed line) FMT approaches. Values of $\kappa_1$ are indicated as labels.} 
\label{zwanzig_compa}
\end{figure}

To explain the linear behaviour obtained for the $\tau$ parameter
as a function of $d$, we have analytically calculated  
the asymptotic behaviour of $\tau$ in the r\'egime $d\gg 1$ for a mixture
of parallel parallelepipeds and cubes in the FMT formulation. 
In Fig. \ref{zwanzig_compa} we compare the $\tau$ parameter, for various
values of $\kappa_1$, obtained from the Zwanzig 
and parallel FMT approaches. An interesting feature is that
depletion is enhanced in the Zwanzig approach, a result
opposite to that from the Onsager approach (Fig. \ref{fig_Zwan}). 

Details on the calculation of the asymptotic behaviour of $\tau$ with 
respect to $d$ have been relegated to Appendix F. 
From the general expression of $\tau$ (Eqn. \ref{ultima}), particularized for the 
parallel case ($Q=1$, which makes the third terms in the numerator and 
denominator to vanish), the linear behaviour can be 
explained if the sum of the two terms in the numerator are of 
order $d^{-2}$. Since the volume ratio is $v_1/v_2=\kappa_1 d^3$, we obtain 
the asymptotic behaviour $\tau\sim\kappa_1 d=L_1/D_2$. This result can also be
obtained using the Zwanzig model (Figs. \ref{tauZwanzig1} and \ref{tauZwanzig}),
which shows that the third term of the numerator in Eqn. (\ref{ultima}) is also of order 
$\sim d^{-2}$. The coefficient $\beta(\kappa_1)$ of the linear fits of $\tau$ with 
respect to $\kappa_1$ (Fig. \ref{tauZwanzig}) 
depends linearly on $\kappa_1$ and has a slope $\beta_2=-0.038$ 
(see caption of Fig. \ref{tauZwanzig}), which is 
in the range of values predicted by the 
asymptotic limit of $W(\kappa_1)\equiv \tau/(\kappa_1 d)$ as a function of 
$\kappa_1$, shown in Fig. (\ref{thelast}).  

The function $\left(\partial\Delta/
\partial\rho_2\right)^{(0)}$ [Eqn. (\ref{com1})] is
of order $d^{-2}$ only within the FMT formulation, the Onsager 
approach giving a term $\sim d^0$. This, in turn,  shows the importance 
of properly taking account of pair correlations between particles in order to 
adequately describe depletion interactions. The
asymptotic limit coincides with the adhesive limit because, when 
the mixture is highly asymmetric, the attractive 
depletion potential between two big particles becomes 
narrower and deeper, tending in the limit to a Dirac delta 
function at contact. This result is confirmed 
by our calculations in Appendix F, 
which show that the inverse Fourier transform of one of the terms of the 
function $\left(\partial\Delta/\partial\rho_2\right)^{(0)}$ 
is a Dirac delta function at contact.
We have repeated the same analysis in Appendix F 
for the case of a HS binary mixture in the FMT formulation (PY approach). 
We have obtained that 
$\left(\partial\Delta/\partial\rho_2\right)^{(0)}_{\hbox{\tiny HS}}\sim {\cal O}(d^{-3})$.
Also, a Dirac delta function at contact 
is present in the inverse Fourier transform of 
$\left(\partial\Delta/\partial\rho_2\right)^{(0)}$.
The difference in the limit of infinite size asymmetry between the HPAR-HC and HS cases
is due to the difference in particle shapes: when two big particles 
are closer enough the parallel sides of 
two parallelepipeds exclude many more small particles from their interstitial 
space than the curved surface of two big hard spheres. The discussion above forces us
to speculate about the possible asymptotic behaviour of the $\tau$ parameter for the
HSPC-HS mixture in the
limit $d\gg 1$: since $v_1/v_2\sim d^3$,  $\tau\sim{\cal O}(d^0)$ 
[Eqns. (\ref{ultima}) and (\ref{com2})]; this is a consequence of the caps of
the spherocylinders being spherical. This argument does not apply to the 
behaviour of $\tau$ in the limit of large $\kappa_1$ for fixed $d$, which continues 
to be linear.

\section{Conclusions and perspectives}
\label{VI}

The first result from this study on depletion effects in smectic phases 
is that, within Onsager theory, the inclusion of orientational degrees 
of freedom significantly reduces the enhanced stability of the smectic phase 
over the nematic phase as compared to that obtained from the
frozen-orientation approximation. A second conclusion that can be drawn 
is that when the surface curvature of the rods is very different from
that of the other species (e.g. the case of HPAR and HS),
depletion is enhanced and the smectic phase stabilizes at lower 
packing fractions as HS become more abundant. This effect is also
reflected in the fact that the N--Sm demixing gap in the phase diagram widens
as one goes from HSPC to HP particles through the HC geometry. 
Finally, the inclusion of more realistic pair correlations between 
particles, in the manner of FMT theory, 
changes the asymptotic behaviour of the $\tau$ parameter (a convenient
measure of depletion in smectics) for large values of the ratio $d$, i.e. 
the asymmetry between the breadth of the rod and the sphere diameter;
in particular, for HPAR, $\tau$ increases linearly with $d$,
while for HSPC
we speculate that $\tau$ goes to a finite limit. The limiting case of large aspect 
ratios and finite $d$ is similarly captured  by both Onsager and Fundamental--Measure
 theories.

In this work we have studied the depletion mechanism along the N--Sm
spinodal line for different mixtures by explicit calculation of the
$\tau$ parameter evaluated on this line in the pure fluid. 
An interesting task could be to extend this study to equilibrium smectic phases,
consisting of well-developed density peaks, by defining some quantity 
evaluated at the smectic density profile. In this case depletion-based mechanisms may add 
interesting phenomenology, such as strong microsegregation between different 
species, which could be directly quantified using some suitably defined
parameter. Work along this direction is currently in progress.

\section*{Acknowledgments}


We are grateful to the referee for his/her comments, suggestions and
careful reading of the manuscript. We thank MIUR (Italy) and Ministerio de
Educaci\'{o}n y Ciencia (Spain) for financial support under the 2005
binational integrated program.
We gratefully acknowledge financial support from Ministerio de
Educaci\'{o}n y Ciencia
 under grants Nos. FIS2005-05243-C02-01, FIS2005-05243-C02-02,
FIS2004-05035-C03-02, BFM2003-0180 and from Comunidad Aut\'onoma de Madrid
(S-0505/ESP-0299).\\YMR was supported by a
Ram\'on y Cajal research contract from the Ministerio de Educaci\'on
y Ciencia.

\renewcommand{\theequation}
{A.\arabic{equation}}
\setcounter{equation}{0}
\section{Appendix}

\section*{Appendix A. Overlap functions:  case of frozen orientations}
\label{Ap1}

When particle orientations are parallel, the functions $f_{ij}(z)$ and
their Fourier transforms can be written exactly for a large class of particle
shapes and their mixtures. 
We consider in turn each of the particle geometries analysed in the
paper.

In the case of mixtures of HSPC
of length $L_1$ and breadth $D_1$, and HS of diameter $D_2$,
the Fourier transforms are
\begin{eqnarray}
\tilde{f}_{11}(k)&=&
\frac{4\pi}{k^3}\left\{\sin\left[k(D_1+L_1)\right]
\right.\nonumber\\\left.\right.\nonumber\\
&-&\left. D_1k\cos\left[k(D_1+L_1)\right]-\sin\left(kL_1\right) 
\right\},\nonumber\\\nonumber\\
\tilde{f}_{12}(k)&=&\frac{4\pi}{k^3}\left\{
\sin\left[(D_{12}+L_1)k/2\right]
\right.\nonumber\\\left.\right.\nonumber\\
&-&\left.\frac{k}{2}D_{12}\cos\left[(D_{12}+L_1)k/2\right]-\sin\left(kL_1/2\right)
\right\},\nonumber\\\nonumber\\
\tilde{f}_{22}(k)&=&
\frac{4\pi}{k^3}\left[\sin\left(kD_2\right)-kD_2
\cos\left(kD_2\right)\right],
\end{eqnarray}
where $D_{12}=D_1+D_2$.

For mixtures of HCYL of length $L_1$ and breadth $D_1$, and HS of diameter $D_2$, the
appropriate Fourier transforms:
\begin{eqnarray}
\tilde{f}_{11}(k)&=& \frac{2\pi D_1^2}{k}\sin\left(kL_1\right),
\nonumber\\&&\nonumber\\
\tilde{f}_{12}(k)&=&2\pi\Bigg\{\frac{D_1^2}{4k}\sin \left[\frac{k(L_1+D_2)}{2}
\right]-
\frac{D_2}{k^2}\cos \left[\frac{k(L_1+D_2)}{2}\right]\nonumber\\ &&\nonumber\\
&+&\frac{2}{k^3}
\left[\sin\left[\frac{k(L_1+D_2)}{2}\right]-\sin\left(\frac{kL_1}{2}\right)
\right]\nonumber\\&&\nonumber\\
&+&\frac{D_1D_2}{2k}\left[\frac{\pi}{2}\left[\cos \left(\frac{kL_1}{2}\right)
J_1\left(\frac{kD_2}{2}\right)\right.\right.\nonumber\\
\left.\left.\right.\right.\nonumber\\
&-&\left.\left.\sin \left(\frac{kL_1}{2}\right)H_1\left(
\frac{kD_2}{2}\right)\right]+\sin\left(\frac{kL_1}{2}\right)\right]\Bigg\}, 
\nonumber\\&&\nonumber\\
\tilde{f}_{22}(k)&=&\frac{4\pi}{k^3}\left[\sin\left(kD_2\right)-kD_2
\cos\left(kD_2\right)\right]
\end{eqnarray}
where $J_1(x)$ and $H_1(x)$ are the Bessel and Struve functions of first order, 
respectively.

For mixtures of HPAR with length and square side length $L_1$, $D_1$, respectively,
and HS of diameter $D_2$, the Fourier transforms are
\begin{eqnarray}
\tilde{f}_{11}(k)&=&\frac{8D_1^2}{k}\sin\left(kL_1\right),
\nonumber\\&&\nonumber\\
\tilde{f}_{12}(k)&=&
8\left\{\left(\frac{D_1}{2}\right)^2\frac{\sin{[k(L_1+D_2)/2]}}{k}\right.\nonumber\\
\left.\right.\nonumber\\
&+&\left.
\frac{\pi}{4}
\left[\frac{2}{k^3}\left(\sin{\left[\frac{k(L_1+D_2)}{2}\right]}-\sin{\frac{kL_1}{2}}\right)
\right.\right.\nonumber\\\left.\left.\right.\right.\nonumber\\
&-&\left.\left.
\frac{D_2}{k^2}\cos{\left[\frac{k(L_1+D_2)}{2}\right]}\right]\right.\nonumber\\
&&\left.\right.\nonumber\\
&\left.+\right.&\left.\frac{D_1D_2}{2k}\left[\frac{\pi}{2}
\left[\cos {\left(\frac{kL_1}{2}\right)}
J_1\left(\frac{kD_2}{2}\right)
\right.\right.\right.\nonumber\\\left.\left.\left.\right.\right.\right.\nonumber\\
&-&\left.\left.\left.
\sin{\left(\frac{kL_1}{2}\right)}
H_1\left(\frac{kD_2}{2}\right)\right]+\sin{\left(\frac{kL_1}{2}\right)}\right]
\right\},
\nonumber\\&&\nonumber\\
\tilde{f}_{22}(k)&=&\frac{4\pi}{k^3}\left[\sin\left(kD_2\right)-kD_2
\cos\left(kD_2\right)\right].
\end{eqnarray}


\renewcommand{\theequation}
{B.\arabic{equation}}
\setcounter{equation}{0}

\section*{Appendix B. Minimisation of Onsager functional in the frozen-orientation 
approximation}
\label{smectic}

We assume a Fourier expansion for the density profiles $\rho_i(z)$:
\begin{eqnarray}
\rho_i(z)\equiv\rho x_i\psi_i(z)=\rho x_i\left(1+\sum_{n=1}^{\infty}s_n^{(i)}
\cos{nkz}\right),
\end{eqnarray}
with $k$ the smectic wave number, and $s_n^{(i)}$ the $n$-th Fourier amplitude for species $i$. 
Introducing this expression into the free-energy funtional, Eqn. (\ref{ONS}), we obtain 
\begin{eqnarray}
\Phi=\rho\Bigg\{\log\rho-1&+&\sum_i x_i\ln x_i+
\frac{x_i}{d_s}\int_0^{d_s}dz\psi_i\left(z\right)\ln \psi_i\left(z\right)
\nonumber\\
&+&\rho {\cal W}(x,k,\{s_k^{(i)}\})\Bigg\},
\end{eqnarray}
Here we have expressed the excess part in terms of the function
\begin{eqnarray}
&&{\cal W}(x,k,\{s_n^{(i)}\})=\frac{1}{2}\sum_{i,j}x_ix_j\nonumber\\
&&\hspace{1cm}\times\left[
\tilde{f}_{ij}(0)+\frac{1}{2}\sum_{n=1}^{\infty} s_n^{(i)}s_n^{(j)}\tilde{f}_{ij}(k_n)\right],
\end{eqnarray}
where $k_n=nk$ and the Fourier transforms $\tilde{f}_{ij}(k)$ were defined in 
Section \ref{II} (since we are assuming perfect orientational order, we set $Q=1$).
From this, the pressure is
\begin{eqnarray}
\beta P=\rho+\rho^2 {\cal W}(x,k,\{s_n^{(i)}\})
\end{eqnarray}
To calculate phase equilibria in the mixture it is more convenient to work in conditions 
of constant pressure. Then, we minimize the Gibbs free energy per particle and
unit thermal energy,
\begin{eqnarray}
g\equiv\frac{1}{\rho}\left(\Phi+\beta P\right)
\end{eqnarray}
with respect to the smectic wave number and the 
coefficients $\{s_n^{(i)}\}$. The 
corresponding derivatives can be written down explicitely but they have to be solved
numerically using an iterative method. In practice, the Fourier expansions 
have been truncated; they include \emph{ca.} 
40 terms so as to satisfy a stringent 
convergence criterion in the iterative procedure (the number of 
Fourier amplitudes was chosen to guarantee an absolute error less than 
$10^{-7}$ in the density profiles). 

\renewcommand{\theequation}
{C.\arabic{equation}}
\setcounter{equation}{0}

\section*{Appendix C. $\tau$ parameter: general case} 
\label{Ap2}
The strategy is to obtain the nematic--smectic spinodal line perturbatively
in the composition $x$, since we only need to know the spinodal in the neighbourhood
of $x=0$ to obtain the $\tau$ parameter.
The equations to solve are three: two defining the spinodal, the third
giving the equilibrium state of the nematic fluid on the spinodal [Eqns. (\ref{ec})].
These equations can be written as
\begin{eqnarray}
&&\Delta(k,\rho_1,\rho_2,Q)=\left[1- \rho_1
\tilde{c}_{11}(k,\rho_1,\rho_2,Q)\right]\nonumber\\\nonumber\\
&&\times\left[1-\rho_2 \tilde{c}_{22}(k,\rho_1,\rho_2,Q)\right]
-\rho_1\rho_2\tilde{c}_{12}^2(k,\rho_1,\rho_2,Q)=0\nonumber\\ 
\label{prima}\\&&\nonumber\\
&&\frac{\partial \Delta(k,\rho_1,\rho_2,Q)}{\partial k}=0
\label{cond}
\\&&\nonumber\\
&&\frac{\partial\Phi(\rho_1,\rho_2,Q)}{\partial Q}=0,
\label{segun}
\end{eqnarray}
where $\tilde{c}_{ij}(k,\rho_1,\rho_2,Q)$ are the Fourier transforms of the 
direct correlation functions between species $i$ and $j$.
The relationship 
between infinitesimal changes in the variables $\{k,\rho_1,\rho_2,Q\}$ 
along the spinodal can be calculated from (\ref{prima}) as 
\begin{eqnarray}
\left(\frac{\partial\Delta}{\partial\rho_1}\right)d\rho_1+
\left(\frac{\partial\Delta}{\partial\rho_2}\right)d\rho_2+
\left(\frac{\partial\Delta}{\partial Q}\right)dQ+
\left(\frac{\partial\Delta}{\partial k}\right)dk=0.\nonumber\\
\label{casi0}
\end{eqnarray}
Similarly, the relation between the changes in the order parameter $Q$ and 
densities $\{\rho_1,\rho_2\}$ can be obtained from (\ref{segun}) as
\begin{eqnarray}
\left(\frac{\partial^2\Phi}{\partial Q\partial\rho_1}\right)d\rho_1+
\left(\frac{\partial^2\Phi}{\partial Q\partial\rho_2}\right)d\rho_2+ 
\left(\frac{\partial^2\Phi}{\partial Q^2}\right)dQ=0.\nonumber\\
\label{casi1}
\end{eqnarray} 
Defining the coefficient $\upsilon\equiv d\rho_1/d\rho_2$, we obtain 
from Eqns. (\ref{casi0}) and (\ref{casi1}):
\begin{eqnarray}
\upsilon=-\frac{\displaystyle{\left(\frac{\partial\Delta}{\partial\rho_2}
\right)-
\left(\frac{\partial\Delta}{\partial Q}\right)
\frac{\left(\partial^2\Phi/\partial Q\partial\rho_2\right)}
{\left(\partial^2\Phi/\partial Q^2\right)}}}{
\displaystyle{\left(\frac{\partial\Delta}{\partial\rho_1}\right)-
\left(\frac{\partial\Delta}{\partial Q}\right)
\frac{\left(\partial^2\Phi/\partial Q\partial \rho_1\right)}
{\left(\partial^2\Phi/\partial Q^2\right)}}},
\label{involve}
\end{eqnarray}
where the condition given in (\ref{cond}) of the absolute minimum of 
$\Delta(k,\rho_1,\rho_2,Q)$ with respect to the wave number $k$ was
used. Note that this expression is valid for any composition $x$ of 
the mixture. Now, evaluating all the derivatives at $\rho_2=0$ we obtain,
from Eqn. (\ref{prima})
\begin{eqnarray}
\left(\frac{\partial\Delta}{\partial\rho_1}\right)^{(0)}
&=&-\rho_0\left[\left(\frac{\partial \tilde{c}_{11}}{\partial\rho_1}\right)
^{(0)}+\frac{\tilde{c}_{11}^{(0)}}{\rho_0}\right],\\
\left(\frac{\partial\Delta}{\partial\rho_2}\right)^{(0)}
&=&-\rho_0\left[\left(\frac{\partial\tilde{c}_{11}}{\partial\rho_2}\right)
^{(0)}+\left(\tilde{c}_{12}^{(0)}\right)^2\right],\label{casi}\\
\left(\frac{\partial\Delta}{\partial Q}\right)^{(0)}&=&
-\rho_0\left(\frac{\partial \tilde{c}_{11}}{\partial Q}\right)^{(0)},
\end{eqnarray}
where the superscripts mean that all derivatives are evaluated 
at $x=0$, $k=k_0$, and $Q=Q_0$, values corresponding to the
one-component fluid spinodal [note that, from Eqn. (\ref{prima}), we get
$1- \rho_0 \tilde{c}_{11}=0$ at $\rho_2=0$, which has been used to obtain 
the derivatives above].

Now, using the fact that, for the one-component fluid, $1-\rho_0\tilde{c}^{(0)}_{11}=0$,
and that the $\tau$ parameter can be expressed through the 
coefficient $\upsilon$ evaluated at $x=0$ as 
$\tau=1+(v_1/v_2)\upsilon^{(0)}$, we finally obtain
\begin{eqnarray}
&&\tau=1-\frac{v_1}{v_2}\nonumber\\\nonumber\\&&\times\frac{\displaystyle{
\left(\tilde{c}_{12}^{(0)}\right)^2+
\left(\frac{\partial \tilde{c}_{11}}{\partial \rho_2}\right)^{(0)}-
\left(\frac{\partial \tilde{c}_{11}}{\partial Q}\right)^{(0)}
\frac{\left(\partial^2\Phi/\partial\rho_2\partial Q\right)^{(0)}}
{\left(\partial^2\Phi/\partial Q^2\right)^{(0)}}}}
{\displaystyle{
\left(\tilde{c}_{11}^{(0)}\right)^2
+\left(\frac{\partial \tilde{c}_{11}}{\partial \rho_1}\right)^{(0)}-
\left(\frac{\partial \tilde{c}_{11}}{\partial Q}\right)^{(0)}
\frac{\left(\partial^2\Phi/\partial\rho_1\partial Q\right)^{(0)}}
{\left(\partial^2\Phi/\partial Q^2\right)^{(0)}}}}\nonumber\\
\label{ultima}
\end{eqnarray}
which is a general expression for the $\tau$ parameter. 
In the Onsager approximation, we have $\tilde{c}_{ij}(k,\rho_1,\rho_2,Q)
=\tilde{f}_{ij}(k,Q)$ and the derivatives of $\tilde{c}_{ij}$ with respect 
to the densities $\rho_i$ vanish, while for parallel rods ($Q=1$) the  
derivative with respect to $Q$ also vanishes, and we obtain 
expression (\ref{tau}). 

For a fluid mixture of particles without orientational degrees of freedom,
such as a mixture of parallel parallelepipeds and cubes or a mixture 
of hard spheres, expression (\ref{ultima}) for $\tau$ can be reinterpreted as follows.
From Eqn. (\ref{ultima}), the $\tau$ parameter can be re-expressed as
\begin{eqnarray}
\tau=1-\frac{v_1}{v_2}\frac{\displaystyle{\left(\frac{\partial S_{ef}^{-1}}
{\partial\rho_2}\right)^{(0)}}}
{\displaystyle{\left(\frac{\partial S_{ef}^{-1}}
{\partial\rho_1}\right)^{(0)}}},
\label{C11}
\end{eqnarray}
where $S^{-1}_{ef}(k,\rho_1,\rho_2)$ is the inverse structure 
factor of an effective one-component fluid of particles, labelled as 
1, with interactions between them being mediated by particles 
labelled as 2. This structure factor can be calculated by evaluating 
the second functional derivative with respect to $\rho_1({\bf r})$ 
of the semi-grand canonical free-energy functional
\begin{eqnarray}
\Upsilon[\rho_1]\equiv F[\rho_1,\rho_2]-\mu_2\int d{\bf r} 
\rho_2({\bf r}),
\label{semi}
\end{eqnarray} 
at the bulk densities\cite{Cuesta}. Here $F[\rho_1,\rho_2]$ is the Helmholtz free-energy 
functional of the binary mixture and the density profile 
$\rho_2({\bf r})$ in Eqn. (\ref{semi}) 
is calculated from the condition of fixed chemical potential of 
species 2: 
\begin{eqnarray}
\frac{\delta F[\rho_1,\rho_2]}{\delta \rho_2({\bf r})}=\mu_2.
\end{eqnarray} 
The result for this effective structure factor is 
\begin{eqnarray}
S_{eff}^{-1}(k,\rho_1,\rho_2)&=&1-\rho_1
\tilde{c}_{11}(k,\rho_1,\rho_2)\nonumber\\
&-&\frac{\rho_1\rho_2\tilde{c}^2_{12}(k,\rho_1,\rho_2)}
{1-\rho_2 \tilde{c}_{22}(k,\rho_1,\rho_2)}.
\end{eqnarray}
After the inclusion of its first derivative with respect to $\rho_1$ and
$\rho_2$, evaluated at $\rho_2=0$ in Eqn. (\ref{C11}), we exactly obtain 
Eqn. (\ref{ultima}) without the third terms in both numerator and denominator
(as they vanish for fluids without orientational degrees of freedom). 

\renewcommand{\theequation}
{D.\arabic{equation}}
\setcounter{equation}{0}

\section*{Appendix D. Correlation functions for HPAR-HC mixture 
in the Zwanzig--Onsager approach} 
\label{corre_zwanzig}

In the Zwanzig approximation the rods point along one of the 
Cartesian axes $x,y$ or $z$. For the binary HP-HC mixture in the Zwanzig approach 
the correlation functions evaluated at ${\bf k}=(0,0,k)$ can be calculated as   
\begin{eqnarray}
-\tilde{c}_{11}(k)&=&2x_{\perp}^2\left[\tilde{f}_{1x,1x}(k)+\tilde{f}_{1x,1y}
(k)\right]+x_{\parallel}^2\tilde{f}_{1z,1z}(k)\nonumber\\
&+&4x_{\perp}x_{\parallel}
\tilde{f}_{1x,1z}(k),\\\nonumber\\
-\tilde{c}_{12}(k)&=&2x_{\perp}\tilde{f}_{1x,2}(k)+x_{\parallel}\tilde{f}
_{1z,2}(k),\\\nonumber\\
-\tilde{c}_{22}(k)&=&\tilde{f}_{22}(k),
\end{eqnarray}
where the subindexes $1\mu$ ($\mu=x,y,z$) label particle 1 (rods),
which point along the direction $\mu$, while the subindex 2 labels cubes. 
$x_{\perp}$ and $x_{\parallel}$ are respectively the fraction of rods 
perpendicular and parallel to the nematic director (which is taken 
to be parallel to the $z$ axis). These variables are functions of the nematic order 
parameter $Q$ [i.e. $x_{\perp}=(1-Q)/3$, and $x_{\parallel}=(1+2Q)/3$], the 
equilibrium values of which should be calculated from the extremum condition  
of the free-energy density with respect to $Q$. This condition reads
\begin{eqnarray}
\frac{\partial\Phi}{\partial Q}=\frac{2}{3}\rho_1\left\{
\ln\left[\frac{1+2Q}{1-Q}\right]-2\rho_1\left(L_1-D_1\right)^2D_1Q\right\}=0
\nonumber\\
\end{eqnarray}
The expressions for the Fourier-transformed overlap functions are 
\begin{eqnarray}
&&\tilde{f}_{1x,1y}(k)=2(L_1+D_1)^2\frac{\sin(kD_1)}{k},\nonumber\\
&&\tilde{f}_{1x,1x}(k)
=8L_1D_1\frac{\sin(kD_1)}{k},\nonumber\\
&&\tilde{f}_{1x,1z}(k)=4D_1(L_1+D_1)\frac{\sin\left[k(L_1+D_1)/2\right]}{k}
,\nonumber\\
&&\tilde{f}_{1z,1z}(k)=8D_1^2\frac{\sin(kL_1)}{k},\nonumber\\
&&\tilde{f}_{1x,2}(k)=2(L_1+D_2)(D_1+D_2)\frac{\sin\left[k(D_1+D_2)/2\right]}{k}
,\nonumber\\
&&\tilde{f}_{22}(k)=8D_2^2\frac{\sin(kD_2)}{k},\nonumber\\
&&\tilde{f}_{1z,2}(k)=2(D_1+D_2)^2\frac{\sin\left[k(L_1+D_2)/2
\right]}{k},
\end{eqnarray}
Note that, due to the discrete axial symmetry of the particles,
some overlap functions can be expressed in terms of others; for instance,
$\tilde{f}_{1x,1x}=\tilde{f}_{1y,1y}$, and 
$\tilde{f}_{1z,1x}=\tilde{f}_{1z,1y}$. 

\renewcommand{\theequation}
{E.\arabic{equation}}
\setcounter{equation}{0}

\section*{Appendix E. Correlation functions 
for HPAR-HC and HS mixtures in the FMT 
approach}

To obtain the Fourier transforms of the correlation 
functions for HP-HC mixtures \cite{Yuri}, and  for HS mixtures \cite{Rosenfeld}, 
we have used the FMT functional, 
which has the same structure in both systems, namely:
\begin{eqnarray}
-\tilde{c}_{ij}({\bf k})&=&\chi^{(0)}\tilde{f}_{ij}({\bf k})+
\boldsymbol{\chi}^{(1)}\tilde{\boldsymbol{R}}_{ij}({\bf k})
+\boldsymbol{\chi}^{(2)}\tilde{\boldsymbol{S}}_{ij}({\bf k})
\nonumber\\ \nonumber\\
 &+&\chi^{(3)}\tilde{ V}_{ij}({\bf k}),
\label{corre_hp}
\end{eqnarray}
where $\tilde{f}_{ij}$, $\tilde{\boldsymbol{R}}_{ij}$, $\tilde{\boldsymbol{S}}_{ij}$ 
and $\tilde{V}_{ij}$ are Fourier transforms of 
the overlap function, and of the mean radius, surface and volume of the 
overlap region between particles $i$ and $j$, respectively. For a HPAR-HC 
mixture the mean radius and the surface area of the overlap region 
are vectors oriented along the directions parallel to the edge lengths (mean radius) 
and perpendicular to the sides (surface area) of the parallelepipeds. 
For the HS mixture they are scalars, as are the
quantities $\chi^{(1)}$ and $\chi^{(2)}$, which for the HPAR-HC mixture are vectors. 
Finally, the correlation functions 
depend on the wave vector ${\bf k}$ or on its module $k$ in the HPAR-HC and HS mixtures, 
respectively. The expressions for $\chi^{(i)}$ 
corresponding to the HPAR-HC mixture are 
\begin{eqnarray}
\chi^{(0)}&=&\frac{1}{1-\eta},\quad \boldsymbol{\chi}^{(1)}=
\frac{\boldsymbol{\xi_2}}
{(1-\eta)^2},\\
\boldsymbol{\chi}^{(2)}&=&\frac{\boldsymbol{\xi_1}}{(1-\eta)^2}
+\frac{2\boldsymbol{\zeta}_2}{(1-\eta)^3},\\
\chi^{(3)}&=&\frac{\rho}{(1-\eta)^2}+\frac{2\boldsymbol{\xi}_1
\boldsymbol{\xi}_2}{(1-\eta)^3}+\frac{6\xi_{2x}\xi_{2y}\xi_{2z}}
{(1-\eta)^4}
\end{eqnarray}
with $\rho=\sum_i\rho_i$ and $\eta=\sum_i\rho_i v_i$ the total density and 
packing fraction, respectively, 
$\boldsymbol{\zeta}_2=(\xi_{2y}\xi_{2z},\xi_{2z}\xi_{2x},
\xi_{2x}\xi_{2y})$, and 
\begin{eqnarray}
\xi_{1x}=\xi_{1y}=\sum_{i}\rho_i D_i,\quad \xi_{1z}=\rho_1 L_1+\rho_2D_2,\\
\xi_{2x}=\xi_{2y}=\rho_1 L_1D_1+\rho_2 D_2^2,\quad \xi_{2z}=
\sum_i\rho_i D_i^2,
\end{eqnarray}
with $L_1$ and $D_1$ the length and breadth of the parallelepiped, while $D_2$ 
is the edge-length of the cube.For the HS mixture we have 
\begin{eqnarray}
\chi^{(0)}&=&\frac{1}{1-\eta},\quad \chi^{(1)}=\frac{\xi_2}{(1-\eta)^2},\\
\chi^{(2)}&=&\frac{\xi_1}{(1-\eta)^2}+\frac{1}{4\pi}\frac{\xi_2^2}
{(1-\eta)^3},\\
\chi^{(3)}&=&\frac{\rho}{(1-\eta)^2}+\frac{2\xi_1\xi_2}
{(1-\eta)^3}+\frac{1}{4\pi}\frac{\xi_2^3}{(1-\eta)^4},
\end{eqnarray}
where 
\begin{eqnarray}
\xi_1=\frac{1}{2}\sum_i\rho_i D_i,\quad \xi_2=\pi\sum_i\rho_i D_i^2,
\end{eqnarray}
and $D_i$ ($i=1,2$) the particle diameters.

To calculate the $\tau$ parameter we only need expressions for 
$\tilde{c}_{11}$ and $\tilde{c}_{12}$. The Fourier transforms of 
the geometric measures of overlapping bodies for the HPAR-HC fluid with 
a wave number ${\bf k}=(0,0,k)$ (smectic symmetry) are
\begin{eqnarray}
\tilde{f}_{11}(k)&=&8D_1^2\frac{\sin (kL_1)}{k},\\
\tilde{\boldsymbol{R}}_{11}(k)&=&4D_1^2\left[D_1\frac{\sin(kL_1)}{k},
D_1\frac{\sin(kL_1)}{k},\right.\nonumber\\
&&\left.4\left(\frac{\sin(kL_1/2)}{k}\right)^2\right],\\
\tilde{\boldsymbol{S}}_{11}(k)&=&2D_1^3\left[4\left(\frac{\sin(kL_1/2)}{k}\right)^2,
4\left(\frac{\sin(kL_1/2)}{k}\right)^2,\right.\nonumber\\
&&\left.D_1\frac{\sin(kL_1)}{k}\right],\\
\tilde{V}_{11}(q)&=&4D_1^4\left(\frac{\sin(kL_1/2)}{k}\right)^2
\end{eqnarray}
The expressions for particles 1 and 2 are 
\begin{eqnarray}
\tilde{f}_{12}(k)&=&2D_{12}^2\frac{\sin(kL_{12}/2)}{k}\\
\tilde{\boldsymbol{R}}_{12}(k)&=&2D_{12}\left[
D_1D_2\frac{\sin(kL_{12}/2)}{k},
D_1D_2\frac{\sin(kL_{12}/2)}{k},\right.\nonumber\\
&&\left.
2D_{12}\frac{\sin(kL_1/2)\sin(kD_2/2)}{k^2}\right],\\
\tilde{\boldsymbol{S}}_{12}(q)&=&2D_1D_2\left[2D_{12}\frac{\sin(kL_1/2)
\sin(kD_2/2)}{k^2},\right.\nonumber\\
&&\left.2D_{12}\frac{\sin(kL_1/2)
\sin(kD_2/2)}{k^2},\right.\nonumber\\
&&\left.D_1D_2\frac{\sin(kL_{12}/2)}{k}\right]\nonumber\\
&&\\
\tilde{V}_{12}(q)&=&4(D_1D_2)^2\frac{\sin(kL_1/2)\sin(kD_2/2)}{k^2},
\end{eqnarray}
where $D_{12}=D_1+D_2$ and $L_{12}=L_1+D_2$.

For HS mixtures these expressions have the form
\begin{eqnarray}
\tilde{f}_{ij}(k)&=&\frac{4\pi}{k^2}\left[\frac{\sin(kD_{ij}/2)}{k}
-\frac{D_{ij}}{2}\cos(kD_{ij}/2)\right],\\\nonumber\\
\tilde{R}_{ij}(k)&=&\frac{4\pi}{k^2}
\left[\frac{\sin(kD_i/2)\sin(kD_j/2)}{k^2}\right.\nonumber\\\left.\right.\nonumber\\
&-&\left.
\frac{D_iD_j}{4}\cos(kD_{ij}/2)\right],\\\nonumber\\
\tilde{S}_{ij}(k)&=&\frac{(4\pi)^2}{k^2}
\left[\frac{D_{ij}}{2}\frac{\sin(kD_i/2)\sin(kD_j/2)}{k^2}\right.\nonumber\\\left.\right.\nonumber\\
&-&\left.
\frac{D_iD_j}{4}\frac{\sin(kD_{ij}/2)}{k}\right],\\\nonumber\\
\tilde{V}_{ij}(k)&=&\frac{(4\pi)^2}{k^4}\left[
\frac{\sin(kD_i/2)\sin(kD_j/2)}{k^2}-\frac{\sin(kD_{ij}/2)}{k}\right.
\nonumber\\\left.\right.\nonumber\\
&+&\left.\cos(kD_i/2)\cos(kD_j/2)\right],
\end{eqnarray}
for $i,j=1,2$.

\renewcommand{\theequation}
{F.\arabic{equation}}
\setcounter{equation}{0}

\section*{Appendix F. Asymptotic behaviour of $\tau$}
\label{asymptotic}

The asymptotic behaviour of $\left(\partial\Delta/\partial\rho_2\right)^{(0)}$ 
for $d=D_1/D_2\gg 1$ can be calculated from (\ref{casi}) and
(\ref{corre_hp}), particularizing the latter equation to HPAR-HC and HS mixtures. 
After some algebra, we arrive at
\begin{eqnarray}
&&\hspace{-0.9cm}
\left(\frac{\partial\Delta}{\partial\rho_2}\right)^{(0)}_{HPAR}=
-\rho_0\left[\left(\tilde{c}_{12}^{(0)}\right)^2+
\left(\frac{\partial\tilde{c}_{11}}{\partial\rho_2}\right)^{(0)}\right]
\nonumber\\\nonumber\\
\sim&-&\frac{v_1}{d^2}y(1+y)\left\{
2(1+y)^2T_1^2(k_0^*/2)\right.\nonumber\\\left.\right.\nonumber\\
&+&\left.\left[T_0(k^*_0)+2yT_1(k_0^*)+y^2T_1^2(k^*_0/2)\right]\kappa_1^{-2}
\right\},
\label{com1}
\end{eqnarray}
where $\kappa_1=L_1/D_1$, $T_0(x)=\cos{x}$, and $T_1(x)=\sin{x}/x$,
while $y=\eta_0/(1-\eta_0)$, and $k^*_0=k_0L_1$. The expression 
corresponding to the HS fluid reads
\begin{eqnarray}
\left(\frac{\partial\Delta}{\partial\rho_2}\right)^{(0)}_{HS}&\sim&
-\frac{v_1}{d^3}y(1+y)\left\{2y^2(1+3y)T_V(k^*_0)+T_{\delta}(k_0^*)
\right.\nonumber\\\left.\right.\nonumber\\&+&\left.
2yT_f(k_0^*)+y(1+6y)T_S(k_0^*)\right\},
\label{com2}
\end{eqnarray}
where we have defined the dimensionless quantities 
$T_V(k_0^*)=6v_1^{-2}\tilde{V}(k^*_0)$, $T_{\delta}(k_0^*)=
3s_1^{-1}\tilde{\delta}(k_0^*)$, $T_f(k_0^*)=3v_1^{-1}\tilde{f}(k_0^*)$, and 
$T_S(k_0^*)=6(v_1s_1)^{-1}\tilde{S}(k_0^*)$, with $s_1$ and $v_1$ the 
surface area and volume of particle 1, while $\tilde{\delta}(q)$ is 
the Fourier transform of the Dirac delta function $\delta(D_1-|{\bf r}|)$.
The presence of a delta function indicates that, in the limit 
$d\to \infty$, the one-component sticky-sphere limit of the fluid is obtained. We arrive at 
the same conclusion, in the same limit, for the HPAR/HC mixture, by 
comparing Eqns. (\ref{com1}) and (\ref{com2}). Each term in 
the right-hand side of Eqn. (\ref{com1}), from left to right,
has the same meaning as in the HS mixture: they are related to the Fourier transform of 
the volume, adhesiveness, overlap function, and surface area of the overlap region 
between two particles, respectively. For example, the inverse Fourier transform of 
$T_0(k^*)$, with the smectic symmetry ${\bf q}=(0,0,q)$, results in  
a term proportional to $\delta(L_1-|z|)/2$, which reflects the stickiness 
of parallelepipeds along the $z$ direction. A complete effective density functional
for the infinite asymmetric limit was worked out for hard-cube mixtures in
Ref.\cite{YuriCuesta}. The direct correlation function resulting from this
functional has a Dirac delta function located at contact of the sides of the cubes.

However, there is an important difference between the HP-HC and HS mixtures, which is 
related to the square and 
cubic power dependence of the expressions (\ref{com1}) and 
(\ref{com2}) with respect to the asymmetric parameter $d=D_1/D_2$. This 
result is related to the fact that the planar geometry of the sides of
parallelepipeds enhances the depletion interaction between two big particles. 
We should compare these results with those obtained using the Onsager approach, 
where $\tilde{c}_{ij}(k,\rho_1,\rho_2)=\tilde{f}_{ij}(k)$ and then 
$\left(\partial \Delta/\partial\rho_2\right)^{(0)}\sim {\cal O}(d^0)$. 
Thus, we can conclude that depletion interactions in this limit cannot be properly
described by the Onsager approximation.

\begin{figure}[h]
{\centering \resizebox{0.45\textwidth}{!}{\includegraphics{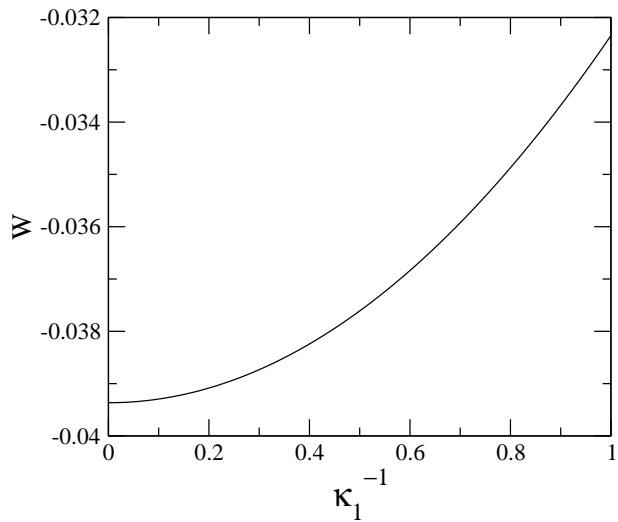}} \par}
\caption{\label{fig_W}\small The function $W(\kappa_1)$.}
\label{thelast}
\end{figure}

Finally, to obtain the asymptotic behaviour of 
the $\tau$ parameter for the HP-HC mixture in the limit $d\to\infty$ , we use 
Eqns. (\ref{ultima}), (\ref{com1}) and (\ref{corre_hp}),
particularized for HP, together with the spinodal instability 
condition $1-\rho_0\tilde{c}_{11}^{(0)}=0$, to obtain explicitly
\begin{eqnarray}
&&\tau\sim \kappa_1 d W(\kappa_1),\\\nonumber\\
&&W(\kappa_1)=-\left\{2(1+y)^2T_1^2(k_0^*/2)+
\left[T_0(k^*_0)+2yT_1(k_0^*)\right.\right.\nonumber\\
\left.\left.\right.\right.\nonumber\\
&&+\left.\left. y^2T_1^2(k^*_0/2)\right]\kappa_1^{-2}\right\}
\left\{y^{-1}\left(3+y^{-1}\right)+2(3+y)T_1(k_0^*)
\right.\nonumber\\\left.\right.\nonumber\\
&&-\left.
\left[3+y+6(1+y)^3\right]T_1^2(k_0^*)\right\}^{-1}
\end{eqnarray}
In Fig. (\ref{fig_W}) the function $W(\kappa_1)$ is plotted. 
As can be seen from the figure, (i) $W$ varies very little
with respect to $\kappa_1$, and (ii) depletion is maximum in
the Onsager limit ($\kappa_1\to\infty$). Note that, on taking
the latter limit, the condition $d=D_1/D_2\gg 1$ must be fulfilled.  


It  is well known that, when the PY approximation is used, the condition 
$1-\rho_0\tilde{c}_{11}(k,\rho_0)> 0$ is always fulfilled 
for all $\eta$ and $q$ in the physical parameter region.  
Because of the absence of a fluid-solid spinodal in this 
approximation, we cannot calculate the $\tau$ parameter to estimate 
the effect of depletion at the freezing transition.  

\end{document}